\documentclass[fleqn,usenatbib]{mnras}
\usepackage{newtxtext,newtxmath}
\usepackage[T1]{fontenc}
\usepackage{ae,aecompl}

\usepackage{multirow}
\usepackage{graphicx}	
\usepackage{amsmath}	
\usepackage{etoolbox}
\usepackage{xcolor}
\usepackage{multirow}
\usepackage{etoolbox}
\def\gtsima{$\; \buildrel > \over \sim \;$}
\def\ltsima{$\; \buildrel < \over \sim \;$}
\def\gsim{\lower.5ex\hbox{\gtsima}}
\def\lsim{\lower.5ex\hbox{\ltsima}}
\def\eeq{\end{equation}}
\def\beq{\begin{equation}}
\def\msun{\hbox{$M_\odot$}}


\title[Redshift Evolution of $M_{\rm H_2}$/$M_{\rm HI}$]{Redshift Evolution of the H$_2$/HI Mass Ratio In Galaxies}

\author[Morselli L. et al.]{
Laura Morselli$^{1,2}$\thanks{E-mail: laura.morselli@unipd.it},
A. Renzini$^{2}$,
A. Enia$^{3,4}$,
G. Rodighiero$^{1,2}$
\\
\\
$^{1}$ Dipartimento di Fisica e Astronomia, Universit{\`a} di Padova, vicolo dell'Osservatorio 3, I-35122 Padova, Italy\\
$^{2}$ INAF $-$ Osservatorio Astrofisico di Padova, vicolo dell'Osservatorio 5, I-35122 Padova, Italy\\
$^{3}$ Dipartimento di Fisica e Astronomia, Universit{\`a} di Bologna, Via Gobetti 93/2, I-40129, Bologna, Italy\\ 
$^{4}$ INAF - Osservatorio di Astrofisica e Scienza dello Spazio, Via Gobetti 93/3, I-40129, Bologna, Italy
}

\date{Accepted 2021 January 18. Received 2021 January 12; in original form 2020 December 11}

\pubyear{2021}

\begin{document}
\label{firstpage}
\pagerange{\pageref{firstpage}--\pageref{lastpage}}
\maketitle

\begin{abstract}

In this paper we present an attempt to estimate the redshift evolution of the molecular to neutral gas mass ratio within galaxies (at fixed stellar mass). For a sample of five nearby grand design spirals located on the Main Sequence (MS) of star forming galaxies, we exploit maps at 500 pc resolution of stellar mass and star formation rate ($M_{\star}$ and SFR). For the same cells, we also have estimates of the neutral ($M_{\rm HI}$) and molecular ($M_{\rm H_2}$) gas masses. To compute the redshift evolution we exploit two relations: {\it i)} one between the molecular-to-neutral mass ratio and the total gas mass ($M_{\rm gas}$), whose scatter shows a strong dependence with the distance from the spatially resolved MS, and {\it ii)} the one between $\log(M_{\rm{H_2}}/M_{\star})$ and $\log(M_{\rm{HI}}/M_{\star})$. For both methods, we ﬁnd that $M_{\rm H_2}$/$M_{\rm HI}$ within the optical radius slightly decreases with redshift, contrary to common expectations of galaxies becoming progressively more dominated by molecular hydrogen at high redshifts. We discuss possible implications of this trend  on our understanding of the internal working of high redshift galaxies.

\end{abstract}




\section{Introduction}
 
Our understanding of galaxy formation and evolution is strictly connected to the accretion of cold gas on galaxies across cosmic time: this gas coming from the cosmic web cools down to form atomic hydrogen (HI) first, and then molecular hydrogen (H$_{\rm 2}$), that can eventually collapse under gravitational instability to form new stars. Feedback from star formation also plays a crucial role, as it is a necessary ingredient to ensure a low efficiency of the star formation process itself: without feedback the gas in a galaxy would be consumed almost completely over a free-fall time, turning most baryons into stars, as opposed to the $\sim 10$ per cent of baryons being locked into stars as actually observed in the local Universe \citep[e.g.][]{2008AJ....136.2846B,2012ApJ...745...69K,2017MNRAS.465.1682H}. Feedback from star formation includes photo-dissociation of H$_{\rm 2}$ into HI due to the radiation emitted by young stars \citep[e.g.][]{2004ApJ...608..314A,2014ApJ...790...10S}. Therefore, HI is not only an intermediate gas phase towards star formation, but also one of its products, and it is key in establishing the self-regulating nature of the star formation process. Unfortunately, till now our knowledge of the HI content in individual galaxies is restricted to the low redshift Universe, where HI is detected in emission via the 21cm line. Several surveys have targeted HI in galaxies at $z<0.05$: HIPASS \citep{2004MNRAS.350.1195M}, ALFALFA \citep{2005AJ....130.2613G}, xGASS \citep{2018MNRAS.476..875C}, HI-MaNGA \citep{masters19}. At higher redshift, the HIGHz survey \citep{catinella_cortese_2015} targeted the HI emission of massive galaxies at $z\sim0.2$, while the CHILES survey pushed the limit of individual detections up to $z\sim0.4$ \citep{fernandez2016}. At even higher redshift our knowledge of HI content is entirely obtained by stacking analysis:  \cite{2016ApJ...818L..28K} at $z\sim1.3$ and \citet[][C20 hereafter]{2020Natur.586..369C} at $z\sim1$. Damped Ly$\alpha$ or MgII absorption line systems give us the chance to estimate the HI content at $z\gtrsim$1.5, with the caveat that they trace HI located well outside the optical disk of galaxies, hence revealing little about what is going on inside their star-forming body. \\ 
\indent Recently, in \citet[][M20 hereafter]{2020MNRAS.496.4606M} we analyzed the HI and H$_{\rm 2}$ content of five nearby, grand-design,  massive main sequence (MS) galaxies on scales of $\sim 500$pc and linked the availability of molecular and neutral hydrogen to the star formation rate (SFR) of each region. We found that H$_2$/HI increases with gas surface density, and at fixed total gas surface density it decreases (increases) for regions with a higher (lower) specific star formation rate (sSFR). In this paper we exploit tight correlations to estimate the evolution with redshift of the H$_2$/HI mass ratio within galaxies. It is generally assumed that this ratio increases with redshift, because galaxies are more gas rich and as the gas surface density increases, recombination is favored. However, galaxies at high redshift  are also more star forming, and higher levels of star formation favor photo-dissociation of the H$_{\rm 2}$ molecule, hence it is not a priori obvious which trend would dominate over the other. 


\section{Data: M$_{*}$, SFR, H$_2$ and HI at 500 \lowercase{pc} Resolution}

The methodology to retrieve estimates of the stellar mass ($M_\star$), SFR, HI mass ($M_{\rm HI}$) and H$_2$ mass ($M_{\rm{H_2}}$) is detailed in \cite{2020MNRAS.493.4107E} and M20. Briefly, starting from the DustPedia archive \citep{2017PASP..129d4102D,2018A&A...609A..37C} we built a sample of five nearby, face-on, grand design spiral galaxies with stellar mass in the range $10^{10.2-10.7}M_{\odot}$, that lie on the MS relation at $z=0$. These sources have been observed in at least 18 bands from the far ultraviolet (FUV) to the far infrared (FIR). We used the photometric data from FUV to FIR to run SED fitting with MAGPHYS \citep{2008MNRAS.388.1595D} on cells of 500pc$\times$500pc. We obtained the SFR as the sum of the un-obscured (SFR$_{\rm UV}$) and obscured (SFR$_{\rm IR}$) contributions. To this aim, SFR$_{\rm UV}$ and SFR$_{\rm IR}$ have been computed using the scaling relations of \cite{bell_kennicutt2001} and \cite{kennicutt98}, respectively, where the UV and IR  luminosities ($L_{\rm UV}$ and $L_{\rm IR}$) are evaluated from the best-fit SED \citep[see][]{2020MNRAS.493.4107E}. Finally, as these sources are included in the HERACLES \citep{2009AJ....137.4670L} and THINGS \citep{2008AJ....136.2563W} surveys, they have been observed in CO(2-1) and HI at 21 cm. Hereafter, we make use of the H$_{\rm 2}$ estimated using $\alpha_{\rm CO}$ = 4.3\msun ${\rm (K\cdot km\cdot s^{-1}pc^2)^{-1}}$ \citep[e.g.][]{2013ARA&A..51..207B}. Details on how the HI and H$_{\rm 2}$ maps at 500pc resolution were obtained can be found in M20,  where the consistency of the results  using a  constant or metallicity-dependent $\alpha_{\rm CO}$ is discussed. 

\section{the H$_2$/HI mass ratio at high redshift}

In this paper we exploit local correlations observed at 500pc resolution to estimate the redshift evolution of the H${\rm_2}$/HI ratio. An important caveat of this procedure is the validity on galactic scales of correlations observed on sub-galactic scales or, in other words, whether integrated quantities can be estimated from spatially-resolved relations. In recent years several studies have indeed revealed that the "main" correlations involved in the star formation process, the MS of star forming galaxies and the molecular gas Main Sequence \citep[MGMS, e.g.][]{2019ApJ...884L..33L} have very similar slopes when analyzed on sub-galactic or galactic scales \citep[e.g.][]{2017ApJ...851L..24H,2019ApJ...884L..33L,2019MNRAS.488.3929C,2020MNRAS.493.4107E}. 

\begin{figure*}
	\includegraphics[width=0.92\textwidth]{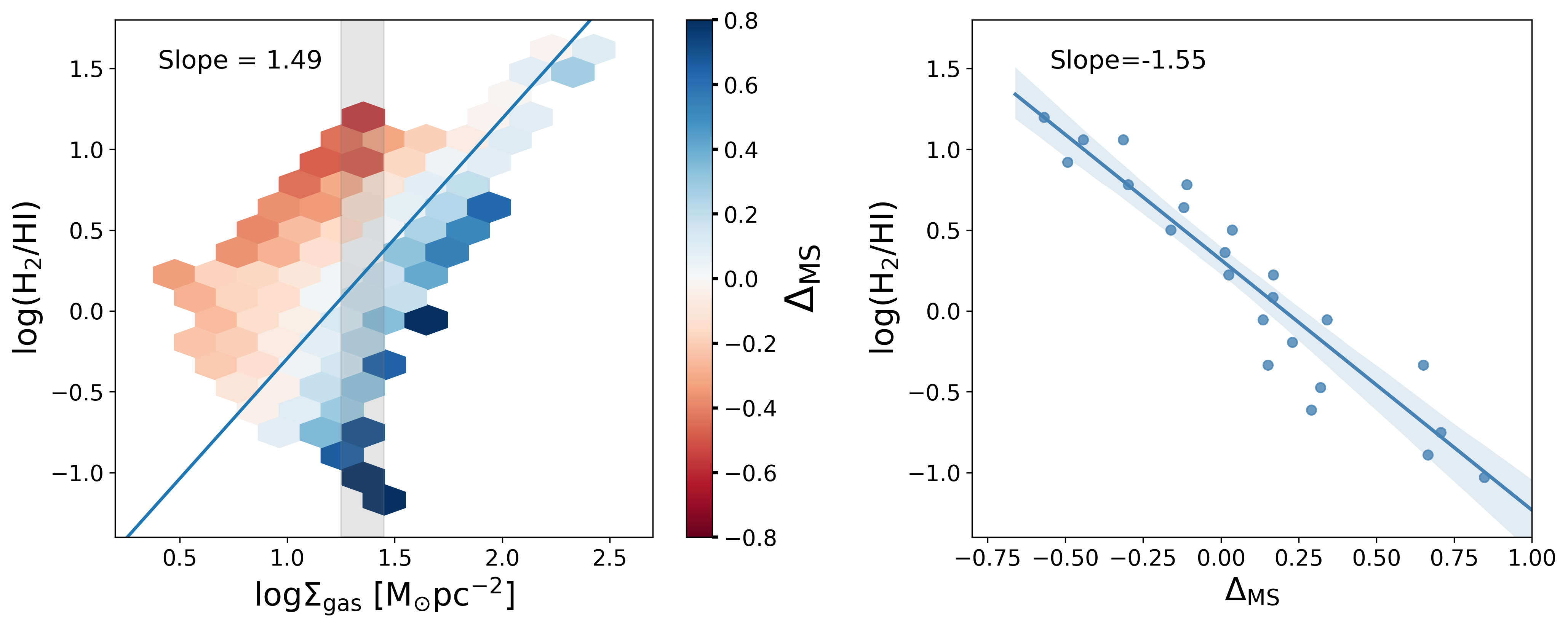}
    \caption{{\it Left panel}: ${\rm log({H_2}/{HI})}$ - ${\rm log}\Sigma_{\rm gas}$ plane, adapted from Figure 8 of M20. Each cell is color-coded according to the average value of $\Delta_{\rm MS}$. The blue solid line is the best fit to the cells having an average value of $\Delta_{\rm {MS}}$ in the range [-0.2,0.2]; the slope of this best fit is $m_1$. The gray shaded area includes the values for which we compute $\Delta_{\rm MS}$ as a function of H$_{\rm 2}$/HI ratio, as shown in the {\it right panel}:  the slope of the best fit (blue solid line) give us $m_2$. }
    \label{fig:fig8}
\end{figure*}

\subsection{Method 1}

To estimate the redshift evolution of $M_{\rm H_2}$/$M_{\rm HI}$ in MS galaxies we proceed as follows. We define the variable $Y$ as the log of $M_{\rm H_2}$/$M_{\rm HI}$ and express it as a function the the total gas mass ($M_{\rm gas} = M_{\rm H_2}+M_{\rm HI}$) and SFR : 
\beq
Y={\rm log}\frac{M_{\rm H_2}}{M_{\rm HI}} = f(M_{\rm gas},{\rm SFR}).
\eeq
\noindent
It follows that:
\beq
{dY\over d{\rm log}(1+z)}=m_1{d{\rm log}M_{\rm gas}\over d{\rm log} (1+z)}+m_2{d{\rm log(SFR)} \over d{\rm log} (1+z)},
\label{eq:3}
\eeq
\noindent
where:
\beq
{\partial Y\over\partial{\rm log}M_{\rm gas}}\simeq m_1 \quad {\rm and} \quad {\partial Y\over \partial{\rm log(SFR)}}\simeq m_2,
\label{eq:m2}
\eeq
with $m_1$ describing the conversion of HI into H$_2$ and $m_2$ the opposite conversion from H$_2$ to HI due to photo-dissociation.
\noindent From \cite{2018ApJ...853..179T} we have that, at fixed stellar mass,
\beq
{d{\rm log}M_{\rm H_2}\over d{\rm log}(1+z)}= 2.6
\label{eq:tacconi}
\eeq
\noindent
which refers only to $M_{\rm{H_2}}$, not to $M_{\rm gas}$. For the redshift evolution of the SFR (at fixed stellar mass) we adopt the scaling from \cite{2014ApJS..214...15S}:
\beq
{d{\rm log }{\rm SFR} \over  d{\rm log} (1+z)} = 3.5.
\label{eq:speagle}
\eeq
\noindent
Therefore, Equation (\ref{eq:3}) becomes:
\beq
{dY\over d{\rm log}(1+z)}=m_1{d{\rm log}M_{\rm gas}\over d{\rm log} (1+z)} +3.5m_2.
\label{eq:9}
\eeq
\noindent
As a next step, we need to derive ${d{\rm log}M_{\rm gas}\over d{\rm log} (1+z)}$. Since we have:
\beq
{\rm log}M_{\rm HI}= {\rm log}M_{\rm H_2}-Y,
\eeq
then:
\beq
M_{\rm gas} = M_{\rm H_2}\times(1+10^{-Y}),
\eeq
\noindent
and the derivative becomes:
\beq
\begin{split}
{d{\rm log}M_{\rm gas}\over d{\rm log}(1+z)} =   {d{\rm log}M_{\rm H_2}\over d{\rm log}(1+z)}+ {d{\rm log}(1+10^{-Y})\over d{\rm log}(1+z)}= \\
2.6 -\left(1+{M_{\rm H_2}\over M_{\rm HI}}\right)^{-1}  {dY\over d{\rm log}(1+z)}
\end{split}
\label{eq:12}
\eeq
where the first derivative is given by Equation (\ref{eq:tacconi}). Therefore, using Equation (\ref{eq:12}), Equation (\ref{eq:9}) becomes:
\beq
\begin{split}
{dY\over d{\rm log}(1+z)}= -m_1\left(1+{M_{\rm H_2}\over M_{\rm HI}}\right)^{-1}  {dY\over d{\rm log}(1+z)} +2.6m_1+ 3.5m_2
\end{split}
\label{eq:13}
\eeq
\noindent Now we integrate the left and right sides of Equation (\ref{eq:13}) between $z=0$ and $z$:
\beq
\begin{split}
& \int_{0}^{z} \left( 1+{m_1\over 1+10^{Y}} \right) \,{dY} = (2.6m_1+3.5m_2)\int_{0}^{\log(1+z)} \,{d{\rm log(}1+z) }
 \end{split}
\label{eq:15}
\eeq
\noindent
By solving the integrals of the left and right sides of Equation (\ref{eq:15}) we get:
\begin{equation}\label{eq:16}
\begin{split}
&(1+m_1)(Y_z-Y_0) + m_1(\log(1+10^{Y_0})-\log(1+10^{Y_z}))\\
& = (2.6m_1+3.5m_2)\log(1+z),
\end{split}
\end{equation}
where the subscript 0 ($z$) refers to the values at redshift 0 ($z$). Thus, this equation is meant to describe the redshift evolution of the H$_2$/HI mass ratio at fixed stellar mass. To proceed with the numerical solution of Equation (\ref{eq:16}), we need the values of $m_1$ and $m_2$ that we obtain from Figure~8 of M20, reported here in the left panel of Figure \ref{fig:fig8}. This Figure shows how the ratio of molecular to atomic hydrogen varies as a function of the total gas surface density and distance from the spatially resolved MS relation, ${\rm \Delta_{MS}}$, which  is defined as the difference between log(SFR) of a region and its MS value at the same stellar mass. Inside galaxies, the H$_{\rm_2}$/HI mass ratio is very strongly correlated with the total gas surface density and anticorrelated with the local SFR, as quantified by ${\rm \Delta_{MS}}$. In M20 we interpret this anticorrelation as evidence that the UV radiation from recently formed, massive stars has the effect of photo-dissociating molecular hydrogen, a manifestation of the self-regulating nature of the star formation process.

We estimate $m_1$ by fitting the relation between ${\rm log({H_2}/{HI})}$ and ${\rm log}\Sigma_{\rm gas}$ along the MS ($\Delta_{\rm{MS}}\sim 0$): the best fit returns a slope of 1.49 (blue solid line in the left panel of Figure \ref{fig:fig8}). To estimate $m_2$, we calculate the slope of the ${\rm log({H_2}/HI)}\!-\!\Delta_{\rm{MS}}$ relation at fixed ${\rm log}\Sigma_{\rm gas}$, considering a narrow range of ${\rm log}\Sigma_{\rm gas}$ values where data exist over the widest range of the ${\rm{H_2}/{HI}}$ mass ratio (the vertical grey region in the left panel), hence offering the best possible estimate of this derivative. The best fit returns a slope of $-1.55$ (right panel of Figure \ref{fig:fig8}). We adopt these two derivatives as proxies for $m_1$ and $m_2$ as defined by Equations (3), based on the aforementioned similarity between the corresponding spatially resolved and global relations. For simplicity, in the following we assume $m_1$=1.5 and $m_2$=$-1.5$, values which are perfectly consistent with the best fit ones. Under this assumption,  Equation (\ref{eq:13}) becomes: 
\beq
{dY\over d{\rm log}(1+z)}=-{1.35\over 1+1.5\left(1+{M_{\rm H_2}\over M_{\rm HI}}\right)^{-1} }\quad.
\label{eq:17}
\eeq
\noindent Equation (\ref{eq:17}) implies that the redshift derivative of $Y$ is always negative, i.e., the phase equilibrium shifts in favour of HI in high redshift galaxies. This comes  from the SFR increasing with redshift faster than the molecular gas mass, see the above Equations (\ref{eq:tacconi}) and (\ref{eq:speagle}). Let us consider three limiting cases. If $M_{\rm H_2}$ largely dominates over $M_{\rm HI}$, then the denominator in Equation (\ref{eq:17}) is $\sim 1$ and the derivative is $-1.35$. If $M_{\rm HI}$ largely dominates, the derivative becomes -0.54. Finally, if the two phases are nearly equal in mass the denominator is $\sim$ 1.75 and the derivative becomes -0.77. So, the derivative will always be between -0.54 and $-1.35$. 

\begin{figure*}
	\includegraphics[width=0.95\textwidth]{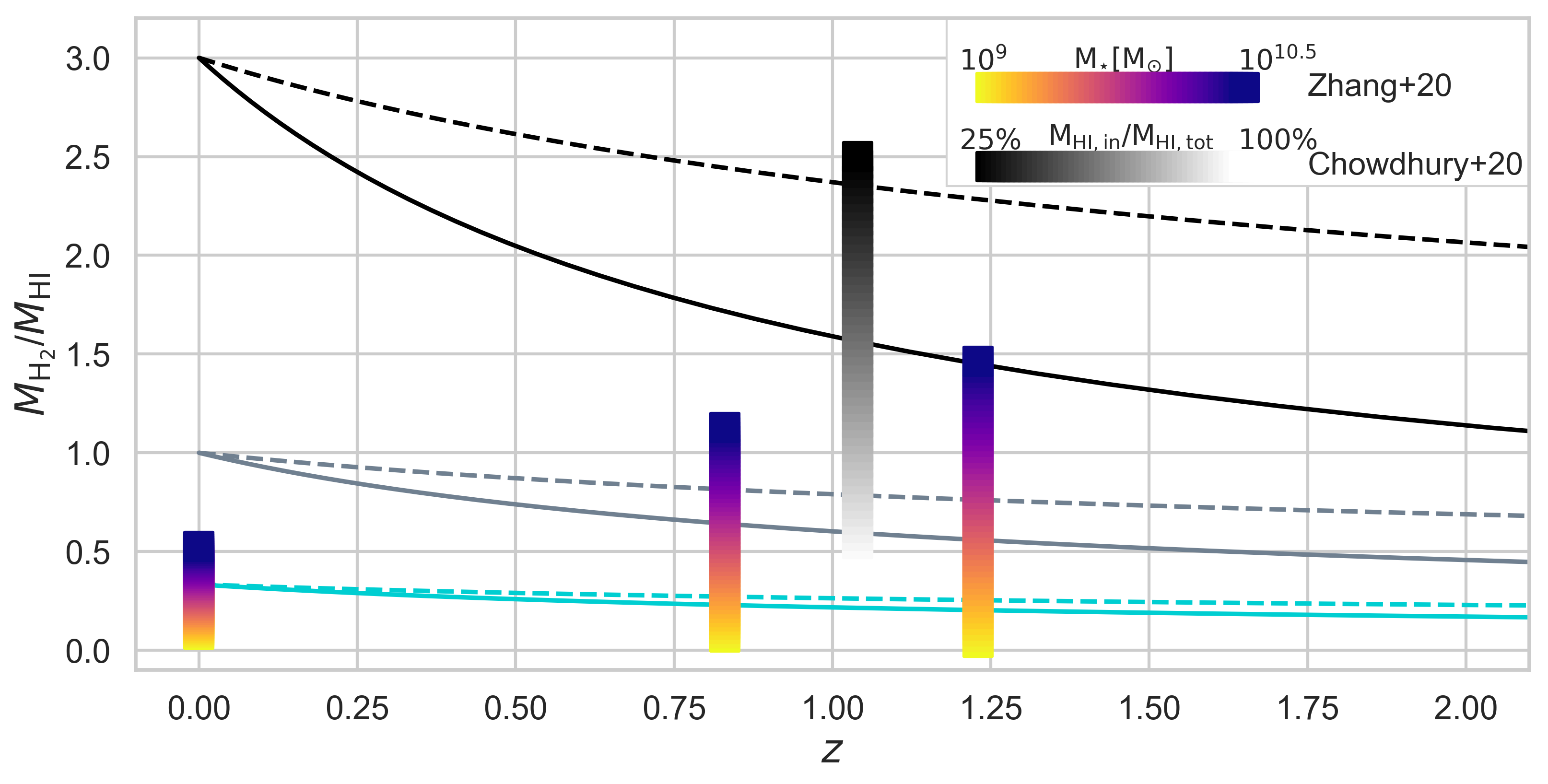}
    \caption{Redshift evolution at fixed stellar mass of the H$_2$/HI mass ratio, obtained applying Method 1 (solid lines) and Method 2 (dashed lines), for three different values of ($M_{\rm H_2}/M_{\rm HI})_{z=0}$= 1/3 (turquoise), 1 (gray) and 3 (black). The values obtain from the HI detection of C20 at z=1.04 are marked with the white-to-black colored bar, with the gradient indicating variations of the fraction of HI inside the optical radius. The values estimated from the correlations of 
   Zhang et al. (2020) at z=0, 0.83 and 1.23 are indicated with the yellow-to-purple colored bar, with the gradient indicating the variations in stellar mass.}
    \label{fig:evolution}
\end{figure*}

However, an analytical solution of Equation~\ref{eq:16} is also possible, 
and it is shown in Figure \ref{fig:evolution} (solid lines) for $m_1=1.5$, $m_2=-1.5$, and for $M_{\rm H_2,0}$/$M_{\rm HI,0}$ = 1/3, 1 and 3, i.e., three typical values of the H$_2$/HI mass ratio within the optical radius of MS galaxies in the local Universe \citep{casasola2020}. Galaxies that at $z=0$ are HI dominated, or in which the two phases are equal in mass, show just a mild evolution of $M_{\rm H_2}$/$M_{\rm HI}$, implying that by $z\sim 2$ HI still holds the majority share. However, galaxies that locally are H$_{\rm 2}$ dominated will tend to show a slightly steeper evolution, to reach $M_{\rm H_2}$/$M_{\rm HI}$ $\sim$ 1.2 at $z=2$. We notice that lower values than 3.5 in Equation~(\ref{eq:speagle}) can be found in the literature: they would imply a flatter evolution of $M_{\rm H_2}$/$M_{\rm HI}$ compared to our results.



\subsection{Method 2}

\begin{figure}
	\includegraphics[width=0.45\textwidth]{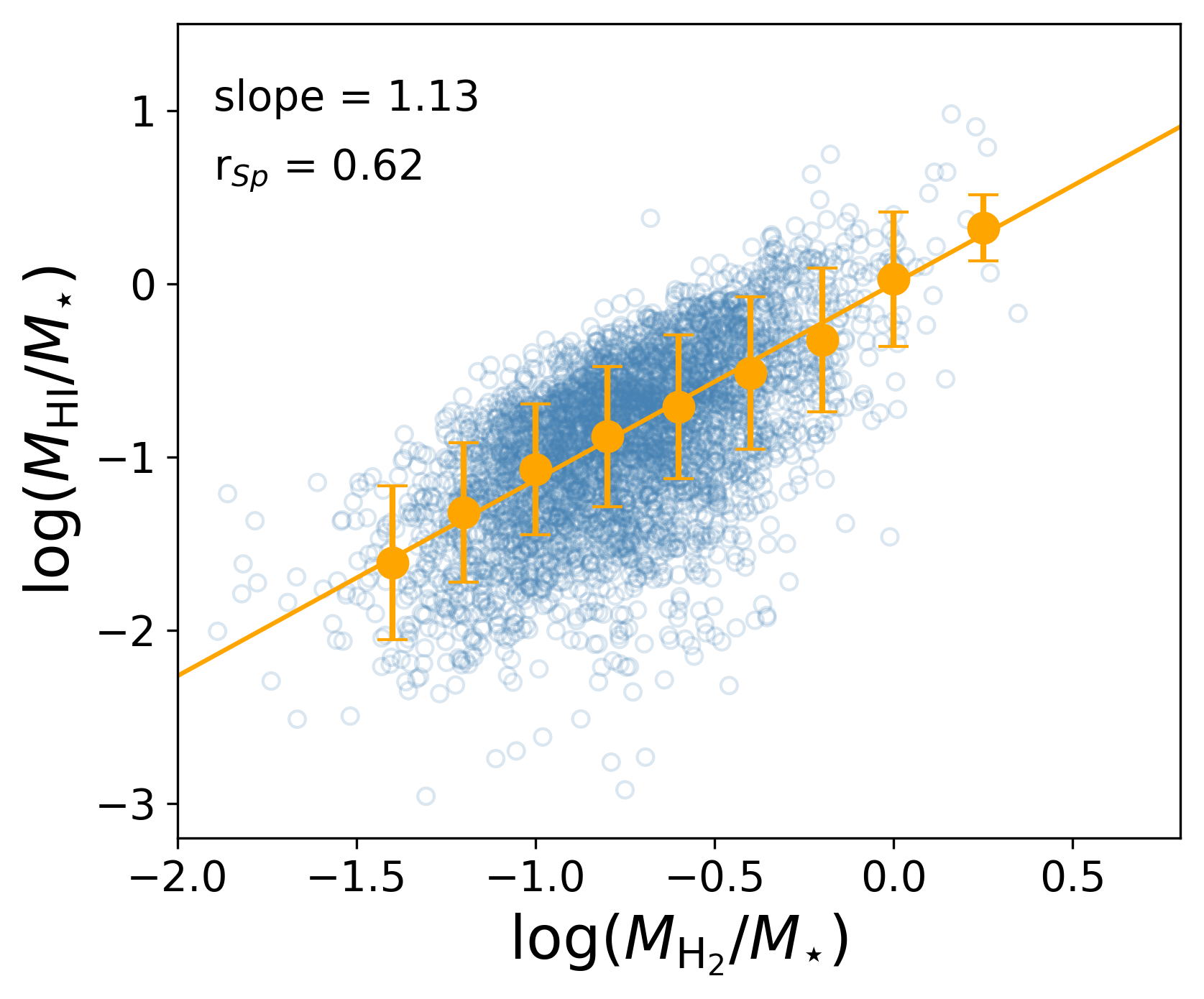}
    \caption{Correlation between log($M_{\rm HI}$/$M_{\star}$) and log($M_{\rm H_2}$/$M_{\star}$) at 500 pc resolution, for the 5 galaxies of M20. The best fit correlation (solid orange line) has a slope of 1.13 and a Spearman coefficient of 0.62.}
    \label{fig:corr}
\end{figure}

With the data for the five galaxies in the sample of M20 we analyze how $M_{\rm HI}$ and $M_{\rm H_2}$ are linked on scales of 500 pc. We observe a slightly super-linear correlation between log($M_{\rm HI}/M_{\star}$) and log($M_{\rm H_2}/M_{\star}$), characterized by a slope of 1.13, a Spearman coefficient of 0.62 and $p$-value $\sim$ 0:
\beq
{\rm log} \frac{M_{\rm HI}}{M_\star} \propto 1.13\ {\rm log} \frac{M_{\rm H_2}}{M_\star} 
\label{eq:hi_from_h2}
\eeq
\noindent and the correlation is shown in Figure \ref{fig:corr}. We note that one of our five galaxies, NGC5194 (M51), has a significantly flatter slope and smaller Spearman coefficient, and interestingly is the only galaxy in the sample to be experiencing an interaction (with M51b) as well as the only one to have T-type = 4 (while the rest of the galaxies have T-type between 5.2 and 5.9). We decided to keep NGC5194 in our sample for consistency with Method 1, but noting that the slope for the remaining four  galaxies is slightly steeper (1.24). This correlation gives us the possibility to estimate the evolution of $M_{\rm HI}/M_\star$ with $z$ just by considering the evolution of the molecular gas (at fixed stellar mass), expressed in Equation (\ref{eq:tacconi}). Hence, Equation (\ref{eq:hi_from_h2}) becomes:
\beq
\frac{M_{\rm HI}}{M_\star} \propto \left(\frac{M_{\rm H_2}}{M_{\star}}\right)^{1.13} \propto (1+z)^{1.13\times2.6}
\label{eq:corr}
\eeq
\noindent
and thus:
\beq
\frac{M_{\rm H_2}}{M_{\rm HI}}\propto (1+z)^{2.6}\times (1+z)^{-2.94}  \propto (1+z)^{-0.34}.
\label{eq:21}
\eeq
\noindent
The trend expressed by Equation (\ref{eq:21}) is shown in Figure ~\ref{fig:evolution} (dashed lines) for the three values of $M_{\rm H_2}$/$M_{\rm HI}$ at $z$ = 0 used in Method 1: 1/3, 1 and 3. The two methods appear to give basically consistent results, with only a modest evolution of $M_{\rm H_2}$/$M_{\rm HI}$ with redshift in favor of HI, which is more pronounced in Method 1 (we note that a steeper slope than the one expressed in Equation (\ref{eq:corr}) would increase the consistency between the two methods). This agreement may not be surprising, as the two methods are in fact more similar than they appear. Indeed, in Method 1 the effect of the SFR on $M_{\rm H_2}$/$M_{\rm HI}$ is treated explicitly, whereas in Method 2 it is implicit in the $M_{\rm H_2}$-$M_{\rm HI}$ correlation. 

\section{Discussion}
\subsection{Comparison with other  estimates of the HI content of high redshift galaxies}
\noindent
We compare these trends with the recent detection of HI in emission in $z\sim1$ galaxies, obtained by C20 via stacking analysis over 7,653 star forming galaxies. They find that in their sample, with a mean stellar mass of $9.4\cdot10^9\,M_\odot$, the mean HI mass is $1.19\cdot 10^{10}\, M_\odot$. To compute the mean H$_2$/HI mass ratio in the galaxies observed by C20 we proceed as follows. We consider the mean molecular-to-stellar mass ratio in the local Universe for galaxies with $M_{\star}\sim10^{10}\,M_\odot$ to be $\sim$ 0.1 \citep[e.g.][]{casasola2020,hunt2020}. Thus, the mean molecular gas mass of galaxies having a mean stellar mass of $9.4\cdot10^9\,M_\odot$ is $\sim 9.4\cdot 10^8\, M_\odot$. By applying the scaling from \cite{2018ApJ...853..179T}, expressed by Equation (\ref{eq:tacconi}), the expected mean $M_{\rm H_2}$ in $z=1$ galaxies turns out to be $\sim 5.7\cdot10^{9}\, M_\odot$, hence:
\beq
\left({\frac{M_{\rm H_2}}{M_{\rm HI}}}\right)_{z=1} =  \frac{5.7\cdot 10^9}{1.19\cdot 10^{10}}=0.48. 
 \label{eq:19}
\eeq
This value is obtained assuming that the HI detected by C20 ($M_{\rm HI_{tot}}$) lies completely within the optical radius ($R_{25}$) of the galaxies in the sample 
(i.e., $f_{\rm R25}=\frac{M_{\rm HI_{R25}}}{M_{\rm HI_{tot}}}=1$, with 
$M_{\rm HI_{R25}}$ the HI mass within $R_{25}$). The average beam of the observations described in C20 is between 30 and 60 kpc, thus it is likely that a certain fraction of the observed HI lies outside the optical radius, resulting in an underestimation of the H$_2$/HI mass ratio within the optical radius. The white-to-black bar in Figure \ref{fig:evolution} represents the estimate of $M_{\rm H_2}/M_{\rm HI}$ at $z$=1, assuming that C20 have sampled a fraction of HI inside the optical radius, varying from 100 per cent (white) to 25 per cent (black). In particular, we find that $M_{\rm HI}>M_{\rm H_2}$ at $z\sim$1 for $f_{\rm R25}>0.4$. For $f_{\rm R25}<0.4$, a value consistent with the $z=0$ estimate of \cite{hunt2020} of galaxies having on average 30$\%$ of their total HI inside the optical radius, we get $M_{\rm H_2}>M_{\rm HI}$ at $z\sim1$, but even when this fraction is only 20\%, $M_{\rm H_2}$ is only a factor of 2 higher than $M_{\rm HI}$.\\
\indent In Figure \ref{fig:evolution} we also include, with yellow-to-purple vertical bars, the recent estimates of $M_{\rm HI}$ obtained by \cite{2020arXiv201104500Z} from the local correlations between
 log${\frac{M_{\rm HI}}{M_\star}}$ and the $(NUV-r)$ color, which is a proxy for the specific SFR. We report their results at three different redshifts: $z$ = 0, 0.83 and 1.23. As above, we use the evolution of $M_{\rm H_2}$ with redshift as given by Equation (\ref{eq:tacconi}) to estimate $M_{\rm H_2}$ at the three redshifts, while $M_{\rm HI}$ is obtained for M$_{\star}$ varying between $10^9\,\msun$ (in yellow in Figure \ref{fig:evolution}) and $10^{10.5}\,\msun$ (in purple in Figure \ref{fig:evolution}). It is worth noting that the HI estimates used in \cite{2020arXiv201104500Z} do not refer to values within the optical radius; this is clear at $z=0$, where the estimates of $M_{\rm H_2}$/$M_{\rm HI}$ are significantly smaller than those of \cite{casasola2020} computed within the optical radius.\\
\indent While our two methods and the one of \cite{2020arXiv201104500Z} yield similar results (in that they suggest a non-vanishing HI contribution at high redshift), it is worth recapping the underlying physical motivations of each of them. Method 1 is built on the observed scaling of the $M_{\rm H_2}/M_\star$ ratio with redshift, Eq. (\ref{eq:tacconi}), and attempts to include the effect of photo-dissociation of the H$_2$ molecules by young stars. Method 2 assumes that the local correlation between $M_{\rm H_2}$ and $M_{\rm HI}$ holds at all redshifts, and the rationale of it is that if galaxies have more H$_2$ they must have also more HI, which is the necessary step to form H$_2$. The method of \cite{2020arXiv201104500Z} assumes that the local correlation between $M_{\rm HI}$ and the ultraviolet-optical colour (a SFR proxy) holds also at all redshifts: as the SFR increases with redshift, so has to do $M_{\rm HI}$ as well. We notice that only our two methods use the observed increasing trend with redshift of the H$_2$ to stellar mass ratio.
\subsection{Implications and conclusions}

All the above results rely on extrapolations from local trends that may or may not hold when applying them to high redshift, thus at this stage we consider the results tentative. Yet, in all methods the H$_2$/HI mass ratio is expected to decrease with redshift, contrary to the notion that it would increase, with H$_2$ dominating at high redshift. Thus, these results suggest that HI cannot be  neglected at high redshift and we discuss below some implications for our understanding of high redshift galaxies.

The first one concerns star formation, namely the gas depletion time $M_{\rm gas}$/SFR and the star formation efficiency (SFE). For lack of direct evidence on the HI mass, the H$_2$ mass has been generally used as a proxy for the total gas mass. If our projections are correct, and if some (if not all) of the HI observed within the optical disk of galaxies comes from H$_2$ photo-dissociation, then the total gas depletion time should be at least a factor of $\sim 2$ longer than previously estimated \citep[e.g.][]{2017ApJ...837..150S,2018ApJ...853..179T}. In the end, the HI to H$_2$ to stars conversion is not a one-way process inside galaxies, but rather a cycle in which part of the H$_2$ in converted back to HI. Thus, the total gas depletion time is a more informative quantity compared to the molecular gas depletion time.

The second implication concerns the contribution of HI to the total baryonic mass inside the stellar disk of high redshift galaxies. Even when ignoring HI, spatially-resolved dynamical studies have shown that $z\sim 2$ galaxies are strongly baryon dominated inside their effective radius \citep{2017Natur.543..397G,2020ApJ...902...98G}. If the mass of HI is comparable to that of H$_2$, as it is at $z\sim 0$, then these galaxies may turn out even more baryon dominated than estimated thus far. Similarly, a higher gas fraction due to the addition of the HI component would lower the Toomre parameter, making disks more prone to clump formation instabilities.

For a direct assessment of the HI content of star forming galaxies at high redshifts we will have to wait for the planned surveys with the Square Kilometer Array (SKA). Indeed, ultra-deep SKA1 surveys may probe massive galaxies (with $M_{\rm HI}\gsim 10^{10}\,\msun$) up to $z\lsim 1.7$ \citep{2015aska.confE.128B},  or even beyond via stacking. Strawman HI surveys with SKA1 foresee two medium/high redshift surveys \citep{2015aska.confE.128B}:  a deep survey (150 deg$^2$) that will detect the mentioned amount of HI up to $z\sim 0.7$ and an ultra-deep survey (2 deg$^2$) that will reach $z\sim 1.7$. These observations should be amply sufficient to check the extent to which our projections are correct.

\section*{acknowledgments}
We are grateful to the anonymous referee for a careful consideration of our manuscript, to Leslie Hunt for useful comments on an early version, and to Lucia Rodr\'iguez-Mu{\~n}oz, Arianna Renzini, Bhaskar Agarwal and Hannah \"Ubler for fruitful discussion and valuable inputs. LM acknowledges support from the BIRD 2018 research grant from the Universit$\grave{\rm a}$ degli Studi di Padova. AE and GR acknowledge the support from grant PRIN MIUR 2017 - 20173ML3WW 001.

\section*{Data Availability}
The derived data underlying this article will be shared on reasonable request to the corresponding author.

\bibliographystyle{mnras}
\bibliography{main}{}

\bsp    
\label{lastpage}
\end{document}